# Identifying the Hidden Nexus between Benford's Law Establishment in Stock Market and Market Efficiency: An Empirical Investigation

*Mostafa Raeisi Sarkandiz* [1]

*Department of Economics, Business and Statistics, University of Palermo, Italy*

**Abstract**

Benford's law, or the law of the first significant digit, has been subjected to numerous studies due to its unique applications in financial fields, especially accounting and auditing. However, studies that addressed the law's establishment in the stock markets generally concluded that stock prices do not comply with the underlying distribution. The present research, emphasizing data randomness as the underlying assumption of Benford's law, has conducted an empirical investigation of the Warsaw Stock Exchange. The outcomes demonstrated that since stock prices are not distributed randomly, the law cannot be held in the stock market. Besides, the Chi-square goodness-of-fit test also supported the obtained results. Moreover, it is discussed that the lack of randomness originated from market inefficiency. In other words, violating the efficient market hypothesis has caused the time series non-randomness and the failure to establish Benford's law.

## 1. Introduction

Benford's law, or the law of the first significant digit, is an empirical rule that explains the distribution of the first digit in a random numerical sequence. Contrary to the common belief that the first digit of the quantities in a numerical sequence appears to adopt a uniform distribution, the evidence display that, in most cases, it follows a logarithmic one (Wang et al., 2009). For the first time, Newcomb (1881) suggested the probability of appearance 1, 2, and 3 as the first digit in natural phenomena is significantly higher than the rest of the six digits. However, his attempt to present a mathematical explanation was unsuccessful. After decades, in a seminal study, Benford (1938) reaffirmed the concept and provided a precise empirical distribution known as Benford's law. The distribution is formulated as follows:

$$P(d_i) = Log\left(1 + \frac{1}{d_i}\right) \quad s.t \quad i \in [1,9] \quad (1)$$

Where $d_i$ represents the first digit of the numbers in a numerical sequence. According to Nigrini & Mittermaier (1997), the classification of Benford's law, in the sense of first digit appearance dispersion, is reported in Table 1.

---

[1] Mostafa.RaeisiSarkandiz@unipa.it

**Table. 1.** *Benford's Law Data Frequency*

| *Digit* | 1 | 2 | 3 | 4 | 5 | 6 | 7 | 8 | 9 |
|---|---|---|---|---|---|---|---|---|---|
| **Probability** | 0.3010 | 0.1761 | 0.1249 | 0.0969 | 0.0792 | 0.0669 | 0.0580 | 0.0512 | 0.0458 |

This exceptional feature has caused Benford's law to be a significant rule in economic fields, particularly accounting and auditing. For instance, if the firms' financial bills do not comply with the law, the possibility of financial statement manipulation will be imminent. For this cause, the U.S. Federal Tax Administration uses this law to recheck the accuracy of financial statements for most U.S. financial institutions (Durtschi et al., 2004).

Several studies have addressed the establishment of law in the stock exchange. For instance, Corazza et al. (2010) investigated the phenomenon by analyzing the daily prices and returns of the S&P 500 index from 1995 to 2007. The outcomes showed that the index did not comply with Benford's law. Moreover, they postulated that the law's disapproval originated from the time series unpredictability and chaotic behavior.

Through a seminal study, Carslaw (1988) indicated that the income time series of active firms on the New Zealand Stock Exchange does not abide by this law. Barbasi et al. (2022) used Benford's Law from the first to kth digit to detect fraud in Italian customs information. The results reject the law's implementation in customs transactions and assess the high probability of data manipulation.

Through another interesting study, Rauch et al. (2011) examined the quality of macroeconomic data related to budget deficit criteria for the EU member states from 1999 to 2009. They came to the conclusion that the macro data for Greece, Romania, Latvia, and Belgium displayed the highest deviation from Benford's law. Furthermore, Ley (1996) concluded if this law is not established in a certain period, by dividing the entire time interval into several sub-periods, its confirmation in some sub-intervals is possible.

On the other hand, using a behavioral perspective, Zgela (2011) pointed out that the law does not apply to data affected by psychological parameters; consequently, the stock price time series will never comply with the law in the long run. In this regard, several studies have pointed out that abnormal trading patterns, such as herding behavior and dead-cat jumps, are clear signs that financial markets are fundamentally governed by psychological parameters (Ko & Huang, 2007).

The present study attempts to identify the underlying reasons for the law violation in the stock market by focusing on the law's chief presumption. The assumption states that data should be distributed randomly. Hence, the randomness property must be checked before testing the phenomenon directly. On the other hand, random distribution of price time series has serious economic consequences. According to Fama's (1970) Efficient Market Hypothesis (EMH), the prices follow a random walk process. Therefore, any attempt to forecast future trends will fail.

Considering the EMH point of view, establishing Benford's law in the stock market means that the time series are distributed randomly, and hence, the market is efficient. In this regard, using a broad range of statistical tests, Raeisi Sarkandiz & Bahlouli (2019) showed that the Warsaw stock market is inefficient. However, the trend can be explained well by Lo's (2004) Adaptive Market Hypothesis (AMH). These findings are supported by other investigations, such as Kilon & Jamroz (2015) and Kompa & Janica (2009).

The novelty of the present study is to shed light on the narrow tie between those two seemingly unrelated concepts employing numerous statistical hypothesis tests. In fact, the study attempts to provide an economic explanation for a purely statistical phenomenon. On the other hand, it will present a new method to test the EMH beyond the routine unit root tests.

The structure of this study is designed as follows. In section 2, the law's theoretical foundations and its statistical aspects are reviewed. In the following, empirical data are introduced in section 3. In section 4, the statistical features are investigated using the non-parametric Run's randomness, the BDS tests, and the parametric test of Chi-square. Moreover, the time series is subjected to unit root tests to examine the EMH. Finally, the concluding remarks are presented in section 5.

## 2. Theoretical Foundations

According to Drake & Nigrini (2000), the law's underlying assumptions are as follows:

1 - The data set must belong to a specific phenomenon that is distributed randomly.

2 - The data set must be complete with no missing data.

3 - The data set must contain significant numerical sequences like addresses or phone numbers.

4 – In the sequence under consideration, the numbers of data with smaller values must be greater than the ones with larger values.

Formann (2010), through a simulation study, showed that sequences with right-unilateral empirical distributions are more likely to comply with Benford's law. Furthermore, the sequences that are obtained through the ratio of two random variables have a higher likelihood of abiding by the law.

Considering the ratio of two similar distributions with different first to four moments. For two uniform distributions, it usually complies with Benford's law. In addition, the ratio of the two exponential distributions possibly follows the law. However, the condition does not hold for Cauchy and normal distributions. For the Chi-square, compatibility with the law depends on the degrees of freedom, such that when the degrees of freedom are equal, the law is well established and decreases with increasing the degree gap. The behavior of the F distribution is quite similar to the Chi-square case. Nevertheless, the speed of deviation of the law is much slower than the Chi-square one (Janvresse & De La Rue, 2004).

The first and second assumptions assert that the data should be distributed randomly with no missing observations. Randomness presumption implies that the data is distributed independently; hence, any long-range correlation is out of the picture. By this definition, randomness rules out the presence of long-memory and mean-reverting features (Baillie, 1996). As a result, a time series that exhibits random characteristics is either stable or has a unit root. On the other hand, stationarity implies the time series is distributed identically but says nothing about whether they are independent (Rahman & Raeisi Sarkandiz, 2023). Nevertheless, a unit root could guarantee that the data is distributed independently.

Mathematically speaking, let $X_t$ be a time series. The most famous process with a unit root is a pure random walk as follows:

$$X_t = X_{t-1} + \varepsilon_t \quad (5)$$

Where $\varepsilon_t \sim IID(0, \sigma^2)$. Suppose $X_0$ is non-zero. Then, the equation can be rewritten as follows:

$$X_t = X_0 + \sum_{i=0}^{t} \varepsilon_{t-i} \qquad (6)$$

Therefore, since the error term is an independent stochastic process, then, a random walk is basically an independent process (Greene, 2002).

From an economic perspective, Fama (1970) introduced the Efficient Market Hypothesis (EMH) and showed that, under some assumptions, it is impossible to forecast future trends. Hence, no one can earn excess returns over the market performance. He postulated that the financial time series follows a random walk process. Therefore, the expected value of future prices is equal to the current one. In other words, the stock market time series has a unit root.

The randomness characteristic is an immediate consequence of such a point of view. Therefore, if the market is efficient, the underlying assumption of Benford's law is satisfied, and thus, there is a possibility for the law establishment. However, numerous empirical and theoretical studies criticized the hypothesis and showed that either the market prices have long memory features or exhibit explosive behaviors such as price bubbles. In this regard, Diba & Grossman (1988) introduced the rational price bubble concept and discussed that most market anomalies originate from the bubbles. Hence, the market cannot be efficient, and time series do not follow a process with a unit root.

Among the empirical studies, investigations of Horvath et al. (2022) and Boubaker et al. (2022) are two recent analyses that empirically discarded the EMH. Consequently, it can be discussed that since the market abides by an inefficient characteristic, violating the randomness assumption closes the door to Benford's law establishment.

According to Drake & Nigrini (2000), in the event of missing data, the likelihood of the law rejection increases. However, excluding the weekends, there are several national holidays, like Independence Day or Easter, when the stock market is closed. Moreover, every four years is a leap year with a 29-day February. Furthermore, there are some unpredicted national day-offs due to weather conditions or similar situations. As a result, looking at the market price time series for a period longer than four years, some unmatched market day-offs could be considered missing data. Hence, from this point of view, one of the fundamental assumptions of Benford's law could be violated, and the possibility of its establishment drops substantially. Nevertheless, unlike randomness, that assumption cannot be tested empirically.

In summary, to examine whether the stock market prices follow Benford's law, one can test it directly through the first-digit distribution employing a Chi-square two-tailed test. Besides, the rejection of the EMH and lack of any unit root can be an indirect method to test the law establishment.

## 3. Data

The present study has used the daily closing prices of the WIG20 index of the Warsaw Stock Exchange from January 1995 to December 2018. The WIG20 indicator is a weighted index of the twenty active large companies of the Warsaw Stock Exchange. The database is restricted to the end of 2018 in order to prevent the turbulence impact of the Covid-19 pandemic on the overall trend. Besides, the data is collected from www.investing.com; hence, the reliability is guaranteed[1].

## 4. Statistical Inferences

In the first step, the descriptive characteristics and the data distribution are analyzed. The normality assumption is investigated by the Jarque & Bera (1980) parametric test. Besides, the uniformity of the empirical distribution is explored using Kolmogorov (1933) and Smirnov (1948) tests. Further, to prevent any mis-conclusion and obtain robust inference, the Anderson & Darling (1952) non-parametric test is utilized. The results are reported in Tables 2 and 3, respectively.

**Table 2.** *Descriptive Statistics*

| *Statistic* | Mean | Median | Std. Dev | Skewness | Kurtosis | Max. | Min. |
|---|---|---|---|---|---|---|---|
| *Value* | 2005.85 | 1952.63 | 664.39 | 0.34 | 2.85 | 3917.87 | 577.90 |

**Table 3.** *Data Distribution*

| *Test* | *Stat.* | *Prob.* |
|---|---|---|
| Jarque- Bera | 117.18 | 0.00 |
| Anderson- Darling | 480.97 | 0.00 |
| Kolmogorov- Smirnov | 0.24 | 0.00 |

The results indicate that neither the normal nor uniform distributions can justify the time series behavior. In the second step, the randomness assumption is analyzed employing Wolfowitz & Wald (1940) Runs and Brock et al. (1996) correlation integral tests, well-known as the Runs test and the BDS, respectively.

The Runs test compares the number of runs in the chief sequence with the number of expected runs and decides if the data is random. So:

$$\bar{R} = \left(\frac{2n_1 n_2}{n_1 + n_2}\right) + 1 \qquad (2)$$

$$S^2 = \frac{2n_1 n_2 (2n_1 n_2 - n_1 - n_2)}{(n_1 + n_2)^2 (n_1 + n_2 - 1)} + 1 \qquad (3)$$

Where $\bar{R}$ represents the expected runs, $S^2$ represents expected runs standard deviation, and $n_1$ and $n_2$ are the number of positive and negative signs in the categorized series, respectively. Now, if $R$ be equal to the real number of runs, the test statistic calculates as follows:

---

[1] https://www.investing.com/indices/wig-20-historical-data

$$Z = \frac{R - \bar{R}}{S} \qquad (4)$$

Under the null hypothesis, the statistic distribution follows the standard normal distribution. Since the test benefits from a two-tailed specification, the rejection of the null does not provide any information about the fundamental trend, and it can be concluded simply that the data are not *I.I.D.*

On the other side, the BDS test is designed based on the correlation integral, which measures the temporal oscillation patterns observed in the data process. Suppose $x_t$ be a time series contain $T$ observations. Then, $x_t^m = (x_t, x_{t-1}, \ldots, x_{t-m+1})$ defines the *m-period history* of the series. The correlation integral with *m-dimensions* will be estimated as follows:

$$C_{m,\varepsilon} = \frac{2}{T_m(T_m - 1)} \sum \sum_{m < s < t \leq T} I(x_t^m, x_s^m; \varepsilon) \qquad (5)$$

Where $T_m = T - m + 1$ and $I(x_t^m, x_s^m; \varepsilon)$ is an Indicator function which defines as follows:

$$I(x_t^m, x_s^m; \varepsilon) = \begin{cases} 1 & \text{if } |x_{t-i} - x_{s-i}| < \varepsilon \text{ for } i = 0, 1, \ldots, m - 1 \\ 0 & \text{otherwise} \end{cases} \qquad (6)$$

The correlation integral estimates the probability of which the distance between two *m-dimensional* points is less than $\varepsilon$. In other words, it approximates the following joint probability:

$$\Pr(|x_t - x_s| < \varepsilon, |x_{t-1} - x_{s-1}| < \varepsilon, \ldots, |x_{t-m+1} - x_{s-m+1}| < \varepsilon) \qquad (7)$$

If $x_t$ be *I.I.D.*, in extreme condition, the above probability will be equals to:

$$C_{1,\varepsilon}^m = \Pr(|x_t - x_s| < \varepsilon)^m \qquad (8)$$

In this regard, Brock et al. (1996) defined the BDS statistic as follows:

$$V_{m,\varepsilon} = \sqrt{T} \frac{C_{m,\varepsilon} - C_{1,\varepsilon}^m}{S_{m,\varepsilon}} \qquad (9)$$

Where $S_{m,\varepsilon}$ represents the standard deviations of $\sqrt{T}(C_{m,\varepsilon} - C_{1,\varepsilon}^m)$. This statistic is convergence in distribution to the standard normal distribution. Therefore:

$$V_{m,\varepsilon} \xrightarrow{D} N(0, 1) \qquad (10)$$

This test for sample size higher than 500, for $m$ lower or equal to 5, and $\varepsilon$ between 0.5 and twice of the standard deviations of data, will be a powerful test. The outcomes of the tests are reported in Tables 4 and 5.

**Table 4.** *Run's Test*

| | |
|---|---|
| *Test Value (Mean)* | 2005.85 |
| *Cases Less Than Test Value* | 2973 |
| *Cases Greater Than Test Value* | 2784 |
| *Total Cases* | 5757   |
| *Number of Runs* | 38  |
| *Z Statistic* | -74.90 |
| *Prob.*       | 0.00   |

**Table 5.** *BDS Test*

| Method: Fraction of Pairs | | Raw Epsilon: 993.1419 | |
| --- | --- | --- | --- |
| *Dimension* | *BDS-Stat.* | *Z-Stat.* | *Prob.* |
| 2 | 0.20 | 239.57 | 0.00 |
| 3 | 0.34 | 257.65 | 0.00 |
| 4 | 0.43 | 279.84 | 0.00 |
| 5 | 0.50 | 311.52 | 0.00 |
| 6 | 0.55 | 354.57 | 0.00 |

In both tests, the null has been rejected at a 5% significant level. Thus, since the underlying assumption of the law is violated, the first digit does not follow the law. Merely, to strengthen the results and substantiate the study's claim through a robustness check, the distribution of the first digit, for both Benford's and uniform distributions, has been examined utilizing the Chi-square statistics. The calculating method of the desired statistic is as follows:

$$\chi^2_{n-1} = \sum_{k=1}^{n} \frac{(O_k - E_k)^2}{E_k} \qquad s.t \quad n \in [1,9] \qquad (11)$$

Where $O_k$ is the number of real observations, and $E_k$ is the number of expected observations. This test was conducted at a 5% significant level. The critical value of the statistic for the 8 degrees of freedom is 15.507. Accordingly, if the estimated value is greater than the critical value, then the null hypothesis will be rejected, and therefore, the series will not conform to the desired distribution. The test results are reported in Tables 6 and 7.

**Table 6.** *Testing a Benford Distribution for First Digit*

| *Digit* | *Expected* | *Actual* | *Expected (%)* | *Actual (%)* | *Deviation* | $\chi^2_{n-1}$ |
| --- | --- | --- | --- | --- | --- | --- |
| 1 | 1732.86 | 2695 | 0.3010 | 0.4681 | 926.14 | 534.21 |
| 2 | 1013.81 | 2396 | 0.1761 | 0.4162 | 1382.19 | 1884.42 |
| 3 | 719.05 | 397 | 0.1249 | 0.0690 | -322.05 | 144.24 |
| 4 | 557.85 | 0 | 0.0969 | 0 | -557.85 | 557.85 |
| 5 | 455.95 | 10 | 0.0792 | 0.0017 | -445.95 | 436.17 |
| 6 | 385.14 | 45 | 0.0669 | 0.0078 | -340.14 | 300.39 |
| 7 | 333.90 | 52 | 0.0580 | 0.0090 | -281.90 | 238 |
| 8 | 294.76 | 139 | 0.0512 | 0.0242 | -155.76 | 82.30 |
| 9 | 263.68 | 23 | 0.0458 | 0.0040 | -240.68 | 219.68 |
| *Sum* | 5757 | 5757 | 100 | 100 | | 4397.26 |

**Table 7.** *Testing a Uniform Distribution for First Digit*

| *Digit* | *Expected* | *Actual* | *Expected (%)* | *Actual (%)* | *Deviation* | $\chi^2_{n-1}$ |
|---|---|---|---|---|---|---|
| 1 | 639.66 | 2695 | 0.1111 | 0.4681 | 2055.34 | 6604.16 |
| 2 | 639.66 | 2396 | 0.1111 | 0.4162 | 1756.34 | 4822.45 |
| 3 | 639.66 | 397 | 0.1111 | 0.0690 | -242.66 | 92.05 |
| 4 | 639.66 | 0 | 0.1111 | 0 | -639.66 | 639.66 |
| 5 | 639.66 | 10 | 0.1111 | 0.0017 | -629.66 | 619.81 |
| 6 | 639.66 | 45 | 0.1111 | 0.0078 | -594.66 | 552.82 |
| 7 | 639.66 | 52 | 0.1111 | 0.0090 | -587.66 | 539.88 |
| 8 | 639.66 | 139 | 0.1111 | 0.0242 | -500.66 | 391.86 |
| 9 | 639.66 | 23 | 0.1111 | 0.0040 | -616.66 | 594.48 |
| *Sum* | 5757 | 5757 | 100 | 100 | | 14857.1 |

As was expected, the hypothesis in both cases has been rejected, and then the distribution of the first digit in the time series neither follows a uniform nor a Benford's distribution.

Finally, the chief claim of this study, which indicates a direct link between market efficiency and the establishment of Benford's law, is investigated using the famous unit root test of Said & Dickey (1984), known as the ADF. The test considers the following specifications:

$$\Delta y_t = \alpha + \beta t + \gamma y_{t-1} + \sum_{i=1}^{p-1} \delta_i \Delta y_{t-i} + \varepsilon_t \qquad (12)$$

Where $\alpha$ is the intercept, $\beta$ stands for the coefficient of deterministic time trend, $\varepsilon_t$ is the $i.i.d$ error terms with variance $\sigma^2$ and zero expected value, and $p$ represents the optimum number of lags which will be selected using information criteria such as AIC or BSC. The model is estimated by the OLS technique. The test inference carries on the estimated value of $\gamma$ coefficient using standard t statistics as follows:

$$DF_\tau = \frac{\hat{\gamma}}{SE(\hat{\gamma})} \qquad s.t. \quad \begin{cases} H_0: \gamma = 0 \\ H_1: \gamma < 0 \end{cases} \qquad (13)$$

It should be noted that the statistic does not follow standard t distribution, and its finite-sample critical values should be obtained from Cheung & Lai (1995). Furthermore, since the test has a left-tailed structure, the rejection of null cannot be interpreted as a covariance stationarity trend. In fact, the rejection of null implies either stationarity or some levels of long-memory characteristics (Raeisi Sarkandiz & Ghayekhloo, 2024). The outcomes are provided in Table 8.

**Table 8.** *ADF Unit Root Test*

| Null Hypothesis: The Time Series Has a Unit Root  Lag Selection: SBC | | |
|---|---|---|
| **ADF-Stat.** | **Critical Value (5%)** | **P-Value** |
| -3.27 | -2.86 | 0.03 |

As can be seen clearly, the presence of a unit root is rejected on a five percent significance level, and hence, the stock prices do not comply with the EMH.

## 5. Conclusion

The present study attempts to link a purely statistical rule of Benford's law to the economic concept of the efficient market hypothesis. To serve the task, the establishment of the law in the stock market has been analyzed. In fact, the paper investigated whether the stock price time series abides by Benford's distribution and what underlying reasons would explain the condition. It has been argued that data randomness is the fundamental assumption for a time series (or any numerical sequence) to follow the law's distribution. On the other hand, the efficient market hypothesis asserts that the market time series complies with a random walk process. As a result, the two concepts are linked through the channel of randomness, such that if the time series are distributed randomly, the market is efficient, and hence, Benford's law would be established.

To support the theoretical discussions, an empirical analysis using the Warsaw stock exchange over an almost long period has been conducted. The outcomes revealed that the time series neither has a unit root nor is distributed randomly. Therefore, it has been concluded that since the market cannot be efficient, according to the Adaptive Market Hypothesis (AMH), then the market time series does not comply with Benford's law, at least in the long horizon.

As explained in the theoretical foundations section, it is possible to find some short periods of time when the market time series obeys the law. In fact, this statement is in line with the AMH's assertion that in some short periods, the market behaves efficiently. Consequently, it is suggested that future investigations focus on the nexus between Benford's law in the stock market and the AMH.

**Funding Sources**

This research did not receive any specific grant from agencies in the public, commercial, or not-for-profit sectors.